\documentclass[english,aps,preprint]{revtex4}
\usepackage[T1]{fontenc}
\usepackage[latin1]{inputenc}
\usepackage{amsmath}
\usepackage{graphicx}
\usepackage{amssymb}

\makeatletter
\newcommand{\p}{\partial}

\makeatother
\begin{document}

\preprint{BROWN-HET-1375}

\title{Real-Time Perturbation Theory in de Sitter Space}

\author{Kevin Goldstein}

\email{kevin@het.brown.edu}

\affiliation{Physics Department, Brown University, Providence, R.I. 02912, USA}

\author{David A. Lowe}

\email{lowe@brown.edu}

\affiliation{Physics Department, Brown University, Providence, R.I. 02912, USA}

\begin{abstract}
We consider scalar field theory in de Sitter space with a general
vacuum invariant under the continuously connected symmetries of the
de Sitter group. We begin by reviewing approaches to define this as
a perturbative quantum field theory. One approach leads to Feynman
diagrams with pinch singularities in the general case, which renders
the theory perturbatively ill-defined. Another approach leads to well-defined
perturbative correlation functions on the imaginary time continuation
of de Sitter space. When continued to real-time, a path integral with
a non-local action generates the time-ordered correlators. Curiously,
observables built out of local products of the fields show no sign
of this non-locality. However once one couples to gravity, we show
acausal effects are unavoidable and presumably make the theory ill-defined.
The Bunch-Davies vacuum state is the unique de Sitter invariant state
that avoids these problems.
\end{abstract}
\maketitle

\newcommand{\bra}[1]{\langle#1|}

\newcommand{\ket}[1]{|#1\rangle}

\section{introduction}

There has been much recent debate about whether quantum field theory
in de Sitter space has a unique vacuum invariant under all the continuously
connected symmetries of the space. The resolution of this question
is crucial to the understanding of possible trans-Planckian effects
on the predictions of inflation \cite{Martin:2000xs}, and observable
effects today such as ultra high energy cosmic ray production \cite{Starobinsky:2002rp,Alberghi:2003br}.
These questions are all the more pressing given recent experimental
results confirming general predictions of inflation for the cosmic
microwave background \cite{Peiris:2003ff}, and of supernova observations
consistent with a positive cosmological constant today \cite{Riess:2000yp}.

At the level of free field theory, de Sitter space has a one-complex
parameter family of vacua, dubbed the $\alpha$-vacua \cite{Chernikov:1968zm,Tagirov:1973vv,Mottola:1985ar,Allen:1985ux}.
It was been argued cutoff versions of these can be relevant during
inflation, where $\alpha$ parameterizes the effects of trans-Planckian
physics \cite{Easther:2001fi,Easther:2001fz,Easther:2002xe,Danielsson:2002kx,Danielsson:2002qh,Goldstein:2002fc,Danielsson:2002mb}.
Others have argued the $\alpha$-vacua suffer from inconsistencies
\cite{Kaloper:2002cs,Einhorn:2002nu,Banks:2002nv,Collins:2003zv}
once interactions are included, and that the Bunch-Davies/Euclidean
vacuum state is the unique consistent state.

In this paper we review existing approaches to this issue, and elaborate
on the connections between them. The most straightforward approach,
where one treats the vacuum state as a squeezed state fails due to
the appearance of pinch singularities, which renders the perturbation
theory ill-defined \cite{Einhorn:2002nu}. We emphasize this is not
a problem with the ultra-violet structure of the theory, but rather
Feynman integrals become ill-defined when propagators on internal
lines are null separated.

A potentially more promising approach based on an imaginary time formulation
\cite{Goldstein:2003ut} leads to a sensible perturbation theory,
and propagators that agree with the imaginary time continuations of
the free propagators of \cite{Mottola:1985ar,Allen:1985ux}. This
perturbative expansion can be continued to real-time and written in
terms of a path integral with a non-local kinetic term, but local
potential and source terms. For the pure scalar field theory, the
algebra of observables built out of local products of the fields remains
local. However once the theory is coupled to gravity the acausality
becomes unavoidable and presumably renders the theory ill-defined,
in keeping with the chronology protection conjecture \cite{Hawking:1992nk}.

\section{Free propagator}

To establish notation, we begin by reviewing the results of \cite{Chernikov:1968zm,Tagirov:1973vv,Mottola:1985ar,Allen:1985ux}
for the free vacua in de Sitter space, invariant under the elements
of the de Sitter group continuously connected to the identity. Fields
may be decomposed as mode sums

\begin{eqnarray*}
\phi(x) & = & \sum_{n}\phi_{n}^{E}(x)a_{n}+\phi_{n}^{E*}(x)a_{n}^{\dagger}\\
 & = & \sum_{n}\phi_{n}^{\alpha}(x)b_{n}+\phi_{n}^{\alpha*}(x)b_{n}^{\dagger}~.\end{eqnarray*}
One then defines the Bunch-Davies/Euclidean vacuum as\[
a_{n}|E\rangle=0\]
and the $\alpha$-vacua as\begin{equation}
b_{n}|\alpha\rangle=0~.\label{eq:alpvac}\end{equation}
The respective mode functions are related as

\[
\phi_{n}^{\alpha}=N_{\alpha}(\phi_{n}^{E}+e^{\alpha}\phi_{n}^{E*})\]

\begin{equation}
\phi_{n}^{E}=N_{\alpha}(\phi_{n}^{\alpha}-e^{\alpha}\phi_{n}^{\alpha*})~,\label{eq:free5}\end{equation}
 with $N_{\alpha}=1/\sqrt{1-\exp(\alpha+\alpha^{*})}$. The creation
and annihilation operators are then related by a mode number independent
Bogoliubov transformation\begin{equation}
b_{n}=N_{\alpha}(a_{n}-e^{\alpha^{*}}a_{n}^{\dagger})~.\label{eq:free5b}\end{equation}
As shown in \cite{Allen:1985ux} we can choose mode functions so that

\begin{equation}
\phi_{n}^{E}(x)^{*}=\phi_{n}^{E}(\bar{x})\label{eq:free6}\end{equation}
with $\bar{x}$ the anti-podal point to $x$. These mode functions
are normalized with respect to the norm\[
(\phi_{1},\phi_{2})=i\int_{\Sigma}(\phi_{1}^{*}\p_{\mu}\phi_{2}-\phi_{2}\p_{\mu}\phi_{1}^{*})d\Sigma^{\mu}\]
with $\Sigma$ any Cauchy surface. 

The Wightman function is

\begin{equation}
\begin{array}{lll}
G^{\alpha}(x,y) & = & \langle\alpha|\phi(x)\phi(y)|\alpha\rangle=\sum_{n}\phi_{n}^{\alpha}(x)\phi_{n}^{\alpha*}(y)\\
 & = & N_{\alpha}^{2}\left(G^{E}(x,y)+e^{\alpha}G^{E}(\bar{x},y)+e^{\alpha^{*}}G^{E}(x,\bar{y})+|e^{\alpha}|^{2}G^{E}(\bar{x},\bar{y})\right)~.\end{array}\label{eq:prop1}\end{equation}
The state $\ket{{\alpha}}$ can be thought of as a squeezed state
with respect to the Euclidean vacuum\[
\ket{{\alpha}}=U\ket{{E}}\]
 with the unitary operator $U$ defined as\[
U=\exp\left(\sum_{n}\beta\left(a_{n}^{E\dagger}\right)^{2}-\beta^{*}\left(a_{n}^{E}\right)^{2}\right)~,\qquad\beta=\frac{1}{4}\left(\log\tanh\frac{-\mathrm{Re}\alpha}{2}\right)e^{-i\mathrm{Im}\alpha}.\]
It is then natural to construct \cite{Goldstein:2003ut}\begin{eqnarray}
\tilde{\phi}(x) & = & U^{\dagger}\phi(x)U\nonumber \\
 & = & N_{\alpha}\sum_{n}\left(\phi_{n}^{E}(x)+e^{\alpha}\phi_{n}^{E}(\bar{x})\right)a_{n}+\left(\phi_{n}^{E}(x)+e^{\alpha}\phi_{n}^{E}(\bar{x})\right)^{*}a_{n}^{\dagger}\label{eq:tphi}\end{eqnarray}
which suggests that from the Euclidean vacuum viewpoint, creation
of a particle in the $\alpha$-vacuum can be thought of as creating
a particle with respect to the Euclidean vacuum at $x$ together with
a particle at the anti-podal point $\bar{x}$.

\subsection{Real-time ordering}

Having discussed the Wightman functions, we now need to discuss more
carefully time-ordering prescriptions. First let us represent de Sitter
space as a hyperboloid in flat $\mathbb{R}^{5}$ with metric $\eta_{ab}=\mathrm{diag}(-1,1,1,1,1)$
and coordinates $X^{a}$ with $a=1\cdots5$\[
X^{a}X^{b}\eta_{ab}=H^{-2}~.\]
Following \cite{Allen:1985ux} we define the signed geodesic distance
between points as

\[
\tilde{d}(x,y)=H^{-1}\arccos\tilde{Z}(x,y)\]
where\[
\tilde{Z}(x,y)=\begin{cases}
{H^{2}\eta_{ab}X^{a}(x)X^{b}(y)+i\epsilon}, & \textrm{{if $x$ to the future of $y$ }}\\
{H^{2}\eta_{ab}X^{a}(x)X^{b}(y)-i\epsilon}, & \textrm{if $x$ to the past of $y$}.\end{cases}\]
With this definition $\tilde{d}(x,y)=-\tilde{d}(\bar{x},\bar{y})$.
Not that although only points with $Z\geq-1$ are connected by geodesics,
$\tilde{d}(x,y)$ can be defined by analytic continuation for $Z<-1$.

The Euclidean vacuum Wightman function is given by\begin{equation}
G^{E}(x,y)=c\;_{2}F_{1}(h_{+},h_{-};2;\frac{1+\tilde{Z}}{2})\label{eq:hypergeo}\end{equation}
where\begin{eqnarray*}
h_{\pm} & \equiv & \frac{3}{2}\pm i\mu\\
\mu & \equiv & \sqrt{{m^{2}-\left(\frac{3H}{2}\right)^{2}}}\\
c & \equiv & \frac{\Gamma(h_{+})\Gamma(h_{-})}{(4\pi)^{2}}~.\end{eqnarray*}
Unless otherwise stated, we consider the case $m>3H/2$ in this paper.
Some of the properties of this function are as follows:

\begin{itemize}
\item a pole when points coincide ($\tilde{Z}=1)$ 
\item a branch cut running along $\tilde{Z}=(1,\infty)$, where the imaginary
part changes sign
\item and asymptotically as $\left|\tilde{Z}\right|\to\infty$\[
G^{E}(x,y)\propto\left(-\tilde{Z}\right)^{-h_{+}}\frac{\Gamma(h_{-}-h_{+})}{\Gamma(h_{-})\Gamma(h_{-}-1)}+\left(-\tilde{Z}\right)^{-h_{-}}\frac{\Gamma(h_{+}-h_{-})}{\Gamma(h_{+})\Gamma(h_{+}-1)}~.\]

\end{itemize}
Using (\ref{eq:prop1}) and (\ref{eq:hypergeo}), the general $\alpha$-vacuum
Wightman function is then explicitly constructed. Note the general
$\alpha$-vacuum Wightman function has poles both at $\tilde{Z}=1$
and $\tilde{Z}=-1$. 

There are a number of options for defining real-time ordering of the
two-point functions described above. Conventional definitions \cite{Allen:1985ux}
correspond to\begin{equation}
iG_{F}^{E}(x,y)=\theta(x^{0}-y^{0})G^{E}(x,y)+\theta(y^{0}-x^{0})G^{E}(y,x)\label{eq:feuc}\end{equation}
and\begin{equation}
iG_{F}^{\alpha}(x,y)=\theta(x^{0}-y^{0})G^{\alpha}(x,y)+\theta(y^{0}-x^{0})G^{\alpha}(y,x)~.\label{eq:feyalp}\end{equation}
with $x^{0}$ a global real-time coordinate. These Green functions
satisfy the inhomogeneous Klein-Gordon equation\[
(\square-m^{2})G_{F}(x,y)=-\frac{\delta^{4}(x-y)}{\sqrt{{-g(x)}}},\qquad g(x)=\det g_{\mu\nu}(x)\]
with $g_{\mu\nu}$ the spacetime metric. Note $\delta(x-\bar{y})$
does not appear. The propagator (\ref{eq:feuc}) can be written as\begin{equation}
iG_{F}^{E}(x,y)=c\;_{2}F_{1}(h_{+},h_{-};2;\frac{1+Z'}{2})\label{eq:fhyper}\end{equation}
where $Z'$ is defined with the new $i\epsilon$ prescription\[
Z'(x,y)=H^{2}\eta_{ab}X^{a}(x)X^{b}(y)+i\epsilon.\]

However, as discussed in \cite{Einhorn:2003xb} another natural time-ordering
in the $\alpha$-vacuum is obtained by ordering the respective terms
of (\ref{eq:tphi}) according to the arguments of the mode functions,
$x$ and $\bar{x}$ (when $\alpha$ is real)\begin{eqnarray}
i\tilde{G}_{EL}^{\alpha}(x,y) & = & N_{\alpha}^{2}\left(\theta(x,y)\left(1+\left|e^{\alpha}\right|^{2}\right)G^{E}(x,y)+\theta(y,x)\left(1+\left|e^{\alpha}\right|^{2}\right)G^{E}(y,x)+\right.\label{eq:einprop}\\
 &  & \left.2\theta(\bar{y},x)e^{\alpha}G^{E}(\bar{y},x)+2\theta(x,\bar{y})e^{\alpha^{*}}G^{E}(x,\bar{y})\right)~.\nonumber \end{eqnarray}

Note these Green functions do not satisfy the asymptotic boundary
condition (\ref{eq:alpvac}) in the infinite past or future. So contrary
to \cite{Einhorn:2003xb} (\ref{eq:einprop}), do not satisfy the
correct boundary conditions for the description of a squeezed state.
These propagators satisfy the non-local inhomogeneous Klein-Gordon
equation \cite{Einhorn:2003xb} \begin{equation}
(\square-m^{2})\tilde{G}_{EL}^{\alpha}(x,y)=-N_{\alpha}^{2}\left(\left(1+\left|e^{\alpha}\right|^{2}\right)\frac{\delta^{4}(x-y)}{\sqrt{{-g(x)}}}+2e^{\alpha}\frac{\delta^{4}(x-\bar{y})}{\sqrt{{-g(x)}}}\right)\label{eq:eldel}\end{equation}
so are to be interpreted as corresponding to source boundary conditions
on a linear combination of podal ($y$) and anti-podal points ($\bar{y}$).

Another natural time ordering is\begin{eqnarray}
i\tilde{G}_{F}^{\alpha}(x,y) & = & N_{\alpha}^{2}\left(\theta(x,y)\left(1+\left|e^{\alpha}\right|^{2}\right)G^{E}(x,y)+\theta(y,x)\left(1+\left|e^{\alpha}\right|^{2}\right)G^{E}(y,x)+\right.\nonumber \\
 &  & \left.\theta(\bar{y},x)\left(e^{\alpha}+e^{\alpha^{*}}\right)G^{E}(\bar{y},x)+\theta(x,\bar{y})\left(e^{\alpha}+e^{\alpha^{*}}\right)G^{E}(x,\bar{y})\right)\nonumber \\
 & = & iN_{\alpha}^{2}\left(\left(1+\left|e^{\alpha}\right|^{2}\right)G_{F}^{E}(x,y)+\left(e^{\alpha}+e^{\alpha^{*}}\right)G_{F}^{E}(x,\bar{y})\right)\label{eq:nfprop}\end{eqnarray}
which is obtained by time ordering the arguments of the propagators
appearing in (\ref{eq:prop1}). This agrees with the propagator of
\cite{Einhorn:2003xb} when $\alpha$ is real, and generalizes it
when $\alpha$ is complex. As we will see, these propagators appear
on internal lines, when one analytically continues from the imaginary
time formulation of \cite{Goldstein:2003ut} to real-time. This propagator
satisfies the inhomogeneous Klein-Gordon equation\[
(\square-m^{2})\tilde{G}_{F}^{\alpha}(x,y)=-N_{\alpha}^{2}\left(\left(1+\left|e^{\alpha}\right|^{2}\right)\frac{\delta^{4}(x-y)}{\sqrt{{-g(x)}}}+\left(e^{\alpha}+e^{\alpha^{*}}\right)\frac{\delta^{4}(x-\bar{y})}{\sqrt{{-g(x)}}}\right)~.\]
This Feynman propagator can be written in terms of hypergeometric
functions using (\ref{eq:fhyper}).

\section{Interacting Theory}

\subsection{Squeezed state approach\label{sub:squeezed}}

The most direct approach to setting up the real-time perturbation
theory is to define Green functions using the interaction picture
representation\begin{equation}
G(x_{1},\cdots,x_{n})=\langle E|U^{\dagger}~T\left(\phi(x_{1})\cdots\phi(x_{n})e^{iS_{int}(\phi)}\right)U|E\rangle\label{eq:squee}\end{equation}
where $S_{int}$ is the interacting part of the action, and $T$ denotes
time-ordering with respect to global time. One can view $U|E\rangle$
as a squeezed state defined in the infinite asymptotic past/future
of de Sitter and these Green functions may then be used to extract
an S-matrix. This expression may be expanded in powers of the interaction
by conventional means, and the end result involves Feynman propagators
ordered with respect to the global time coordinate (\ref{eq:feyalp}).

As emphasized in \cite{Einhorn:2002nu}, pinch singularities arise
in these expressions which render the integrals ill-defined. This
may be seen in the following example of a diagram that occurs with
a $\lambda\phi^{4}$ interaction. The lower loop gives rise to the
factor

\begin{figure}
\includegraphics[%
  clip]{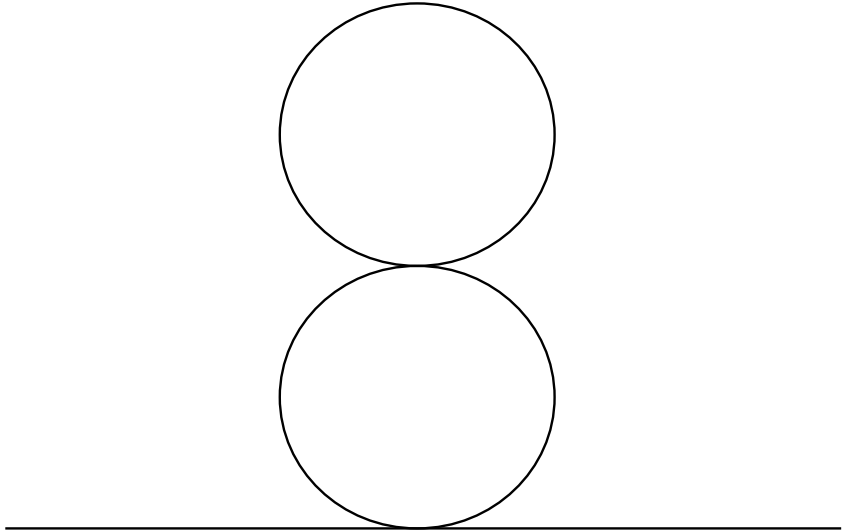}

\caption{Pinch diagram\label{cap:Pinch-diagram}}
\end{figure}
\begin{equation}
\int_{dS_{4}}d^{4}x\sqrt{{-g}}G_{F}^{\alpha}(x,y)^{2}\label{eq:collins}\end{equation}
which contains terms like\[
\int_{dS_{4}}d^{4}x\sqrt{{-g}}G_{F}^{E}(x,y)G_{F}^{E}(\bar{x},\bar{y})~.\]
Writing this in terms of hypergeometric functions (\ref{eq:hypergeo}),
we see the $i\epsilon$ prescriptions differ in the two propagators.
The contribution coming from the region of integration where $x$
is null related to $y$ gives a term proportional to\begin{equation}
\int dZ\frac{1}{Z-1+i\epsilon}\frac{1}{Z-1-i\epsilon}=\frac{\pi}{\epsilon}\label{eq:pinch}\end{equation}
where $Z$ is integrated along the real axis. As $\epsilon\to0$ the
poles pinch the integration contour and the integral diverges. We
emphasize this divergence has nothing to do with the ultraviolet structure
of the theory, so cannot be regulated with local (or even bi-local)
counter-terms. 

This divergence implies that perturbation theory does not make sense
as it stands. It is conceivable that some resummation of perturbation
theory does make sense, but we lack methods to address this kind of
approach in a completely systematic way. We note this type of resummation
is attempted in non-equilibrium statistical field theory where one
similarly encounters pinch singularities \cite{Altherr:1994fx,Altherr:1995jc}.
This resummation can lead to a shifting in the poles of the propagator,
so that the $i\epsilon$ in (\ref{eq:pinch}) is replaced by an $i\Gamma$
where $\Gamma$ is a decay rate, proportional to some power of the
interactions. It is then clear from (\ref{eq:pinch}) that the resummed
theory will be difficult to handle due to the appear of inverse powers
of the coupling.

\subsection{Imaginary-time approach}

Since the direct approach to treating the $\alpha$-vacua as squeezed
states in de Sitter space is a nonstarter, one can try to appeal to
an imaginary time formulation \cite{Goldstein:2003ut}. This provides
us with a straightforward way to deal with spacetimes with event horizons,
since for imaginary time the event horizon shrinks to a point. In
the black hole case, field theory on the imaginary time continuation
(also called the Euclidean section) of a black hole background leads
to a density matrix description from the real time point of view.
One might have hoped a similar novel interpretation of the $\alpha$-vacua
emerges in the real-time point of view, due to the cosmological horizon
of de Sitter space.

To discuss the continuation from real time to imaginary time we use
global coordinates

\[
ds^{2}=-dt^{2}+\left(\cosh t\right)^{2}d\Omega_{3}^{2}\]
so $t\to i\tau$ takes us to imaginary time. Here we have chosen units
where $H=1$. The imaginary time continuation of de Sitter is the
four-sphere. The imaginary time approach of \cite{Goldstein:2003ut}
proceeds by using the transformation (\ref{eq:tphi}) to express $\alpha$-vacuum
Green functions as linear combinations of Euclidean vacuum Green functions\begin{equation}
G(x_{1},\cdots,x_{n})=\langle E|U^{\dagger}\phi(x_{1})\cdots\phi(x_{n})e^{iS_{int}(\phi)}U|E\rangle=\langle E|\widetilde{\phi}(x_{1})\cdots\widetilde{\phi}(x_{n})e^{iS_{int}(\tilde{\phi)}}|E\rangle~.\label{eq:pfac}\end{equation}
This is to be understood as an interaction picture expression. An
important point is that on the Euclidean section, the free propagators
$G^{\alpha}(x,y)$ are symmetric functions of their arguments because
points are spacelike separated. The ordering of operators in this
expression is therefore irrelevant. As described in \cite{Goldstein:2003ut}
the ultraviolet divergences that appear in this approach can be canceled
by de Sitter invariant local counter-terms. This approach yields free
two-point functions that match those of Mottola and Allen on the Euclidean
section, and it is in this sense the approach is a generalization
of the $\alpha$-vacuum to the interacting case. However the real-time
physics of this approach is so far mysterious, and we wish to explore
this question in the following.

\subsection{Continuation to real time}

\begin{figure}
\includegraphics{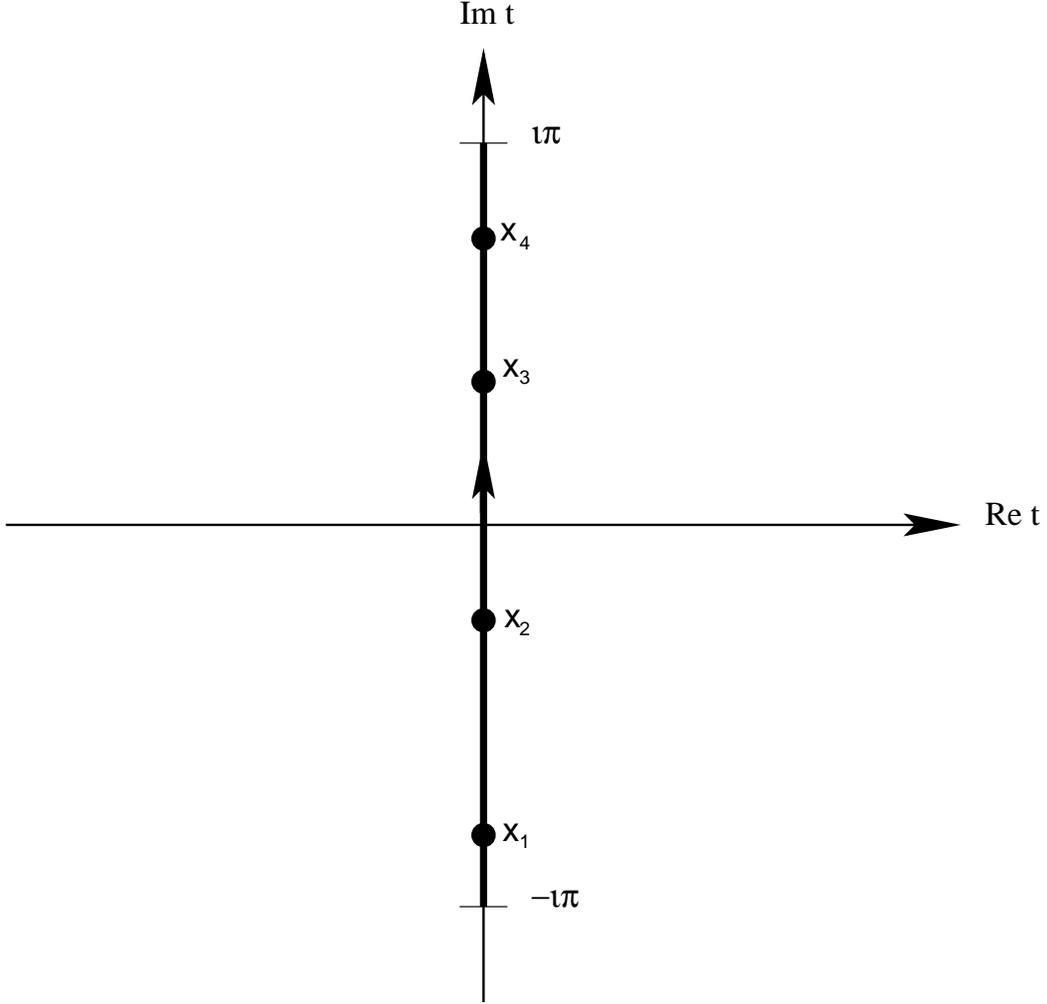}

\caption{Imaginary time contour.\label{cap:Imaginary-time-contour.}}
\end{figure}

\begin{figure}
\includegraphics{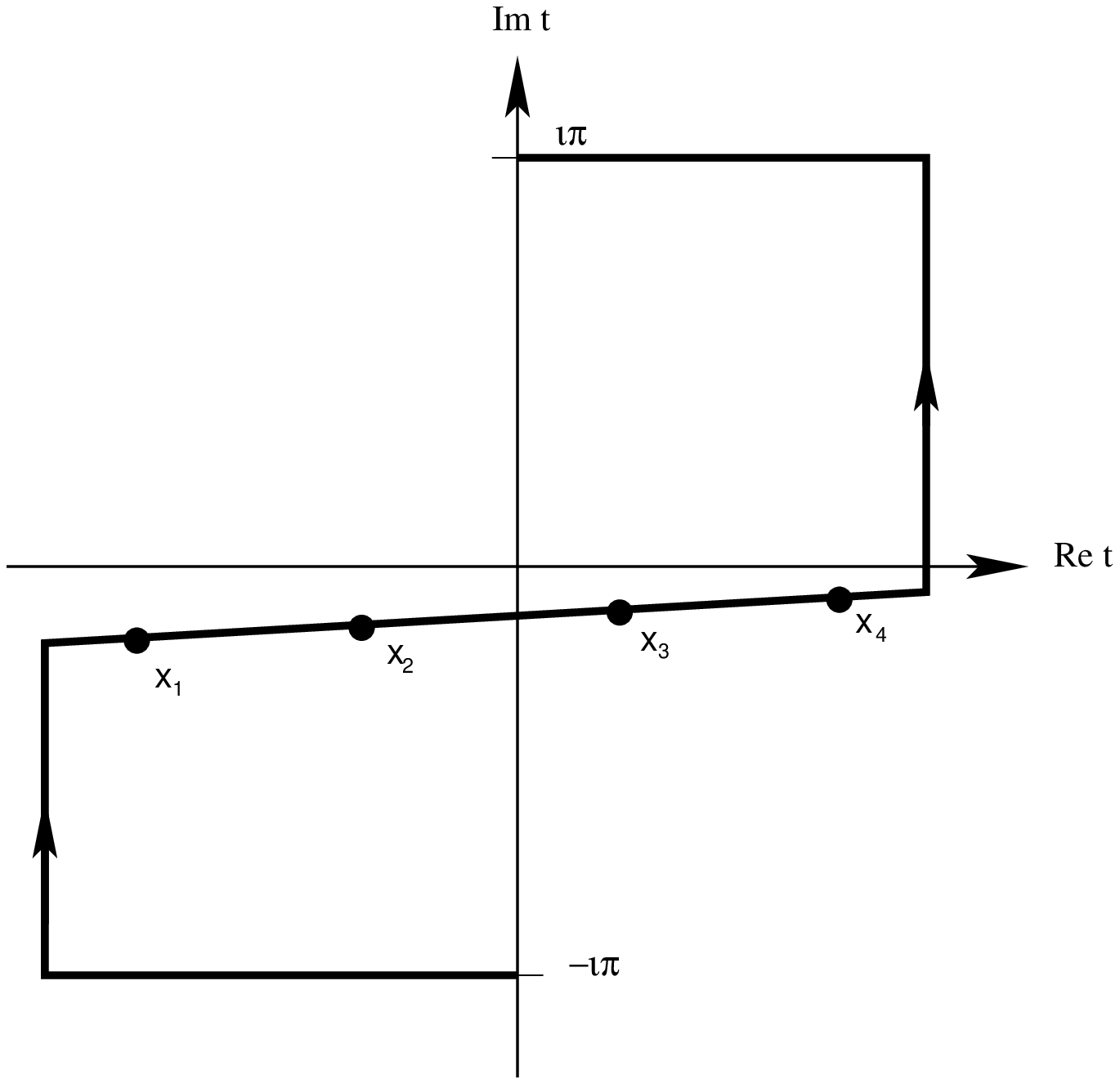}

\caption{Real-time contour.\label{cap:Real-time-contour.} The horizontal
components of contours are to be understood to run off to $t=\pm\infty$.}
\end{figure}

Let us examine what happens when we continue integrals of products
of $G^{\alpha}(x,y)$ on the sphere to integrals over de Sitter space.
We are free to expand (\ref{eq:pfac}) as in (\ref{eq:prop1}) and
order the arguments in any way convenient for analytically continuing
to real-time. Let us first focus on the case of the Euclidean vacuum
$e^{\alpha}=0$. We begin with the imaginary time contour as shown
in figure \ref{cap:Imaginary-time-contour.} running from $-i\pi$
to $i\pi$. This may be continued to the contour shown in figure \ref{cap:Real-time-contour.}.
The horizontal component (with a small positive slope) running from
$-\infty$ to $\infty$ through points $x_{1}$ to $x_{4}$ gives
rise to the expected real-time contour. The corresponding $i\epsilon$
prescription is $t\to t+i\epsilon~\mathrm{sgn}~t$ which gives a Feynman
propagator connecting internal lines (\ref{eq:fhyper}). The vertical
components of the contour are closely analogous to those that appear
in the real-time formulation of finite temperature field theory in
flat space \cite{Landsman:1987uw,Niemi:1984nf}. There the vertical
components of the contour typically factorize for Green functions
evaluated at finite values of the time. However this factorization
is quite subtle \cite{Evans:1995gb} even for flat space, so we will
not assume it here in general. 

The other horizontal components to the time-contour correspond to
the fact that one is not computing a transition amplitude, but rather
the expectation value of some time-ordered string of field operators
with respect to a density matrix specified at some specific time $t_{0}$
(where $t_{0}=0$ in figure \ref{cap:Real-time-contour.})\begin{equation}
G(x_{1},\cdots,x_{n})=\mathrm{Tr}\rho(t_{0})T\left(\phi(x_{1})\cdots\phi(x_{n})\right)~.\label{eq:denmat}\end{equation}
The additional time contour represents the time-evolution back to
the initial time, as is easy to see when (\ref{eq:denmat}) is written
in Schroedinger picture. The density matrix formalism is discussed
in the general context of Friedman-Robertson-Walker backgrounds in
\cite{Semenoff:1985kr}. In this work a deformation of the contour
shown in figure \ref{cap:Real-time-contour.} is used that only contains
two horizontal real-time components (and a single vertical component). 

In \cite{Collins:2003zv} the same two-component time-contour is used
(neglecting the vertical components of the contour), which can be
re-expressed in terms of a two-component field formalism. However
they use a time-ordering prescription analogous to (\ref{eq:feyalp}),
which as we will see is not obtained via analytic continuation of
the imaginary time approach. Their main claim was that there exist
ultra-violet divergences that cannot be canceled with de Sitter invariant
counter-terms. They approximated the $\alpha$-vacuum by starting
with the free field theory state, and then turned on interactions
at a finite time. Given that their boundary conditions did not respect
de Sitter invariance, the appearance of de Sitter non-invariant counter-terms
should not be too surprising. Such features cannot arise in a manifestly
de Sitter invariant formulation, so are more properly regarded as
renormalization effects associated with the symmetry breaking initial
state. 

It is natural to conjecture that in-out transition amplitudes relating
the vacuum in the infinite past to vacuum in the infinite future,
may be computed by including only the horizontal component of the
contour running along the real time axis (with small positive slope)
as shown in figure \ref{cap:Real-time-contour.}. This allows us to
express the real-time physics using a single component field. This
fits nicely with the expectation that pure states do not evolve into
mixed states in a fixed de Sitter background, since the spacetime
is globally hyperbolic %
\footnote{In general, when continuing from imaginary-time Green functions to
real-time Green functions, pure states evolve to mixed states \cite{Hawking:1982dj}.%
}.

Now let us examine what happens when $e^{\alpha}\neq0$. Again we
start with the imaginary time contour, with a product of imaginary
time propagators (\ref{eq:prop1}). These may be decomposed into Euclidean
vacuum correlators as in the second line of (\ref{eq:prop1}). To
continue to real-time, we continue to the contour shown in figure
\ref{cap:Real-time-contour.}. The real-time continuation of the imaginary
time formalism \cite{Goldstein:2003ut} therefore yields a set of
amplitudes of the form (\ref{eq:denmat}), with $\phi$'s replaced
by $\tilde{\phi}$'s.

If we wish to compute only in-out matrix elements, then we retain
only the horizontal contour running along the real axis, and we find
podal and anti-podal points are to be ordered according to their global
time $t$ \cite{Einhorn:2003xb}. Thus the propagators (\ref{eq:nfprop})
will appear on internal lines in the general $\alpha$-vacuum expression
for real-time ordered Green functions. Because all singularities appearing
in the propagator have the same $i\epsilon$ prescription, the resulting
integrals are integrals of analytic functions, and no pinch singularities
arise. Likewise, no pinch singularities will arise if we use the full
contour to compute the continuation of the amplitudes of \cite{Goldstein:2003ut}
, though one must then use a multi-component field formalism analogous
to \cite{Semenoff:1985kr} to directly perform the real-time computations.

One might wonder whether the single horizontal component plus the
vertical components of the contour might lead one back to the in-out
amplitudes of the squeezed state approach. This cannot be the case,
because the vertical components will not generate pinch singularities,
nor will the vertical components change the boundary conditions on
the free propagators from (\ref{eq:feyalp}) to (\ref{eq:nfprop}).

\subsection{Path Integral Formulation}

The real-time continuation of the imaginary time formalism just described
yields a set of finite renormalized in-out amplitudes of the form\begin{equation}
G(x_{1},\cdots,x_{n})=\langle E|T\left(U^{\dagger}\phi(x_{1})\cdots\phi(x_{n})e^{iS_{int}(\phi)}U\right)|E\rangle=\langle E|T\left(\widetilde{\phi}(x_{1})\cdots\widetilde{\phi}(x_{n})e^{iS_{int}(\tilde{\phi)}}\right)|E\rangle\label{eq:tord}\end{equation}
where the time-ordering prescription is as in (\ref{eq:nfprop}).
Note that by definition local sources couple to the $\tilde{\phi}$
\cite{Goldstein:2003ut}, which is a linear combination of the field
at podal and anti-podal points as in (\ref{eq:tphi}). We will work
under the hypothesis that the set of amplitudes (\ref{eq:tord}) define
a consistent set of probability amplitudes. For example, these could
be used to approximate transition amplitudes corresponding to observations
of a comoving observer. The general set of local observables should
correspond to Wightman functions of the $\tilde{\phi}$. 

The time-ordered correlators (\ref{eq:tord}) can be generated from
the following path integral\[
Z=\int\mathcal{D}\tilde{\phi}e^{iS[\tilde{\phi}]}\tilde{\phi}(x_{1})\cdots\tilde{\phi}(x_{n})\]
where the action $S$ is\begin{equation}
S=\frac{1}{2}\int d^{4}x\sqrt{{-g(x)}}~d^{4}y\sqrt{{-g(y)}}\tilde{\phi}(x)K(x,y)\tilde{\phi}(y)-\int d^{4}x\sqrt{{-g(x)}}\left(V(\tilde{\phi})+j(x)\tilde{\phi}(x)\right)\label{eq:locac}\end{equation}
with the non-local kinetic term determined by\begin{eqnarray}
K(x,y) & = & \left(a\frac{\delta^{4}(x-y)}{\sqrt{{-g(x)}}}+b\frac{\delta^{4}(x-\bar{y})}{\sqrt{{-g(x)}}}\right)\left(\square_{x}-m^{2}\right)\label{eq:kinetic}\\
a & = & \frac{1-|e^{\alpha}|^{4}}{(1-e^{2\alpha})(1-e^{2\alpha^{*}})}\nonumber \\
b & = & -\frac{(e^{\alpha}+e^{\alpha^{*}})(1-|e^{\alpha}|^{2})}{(1-e^{2\alpha})(1-e^{2\alpha^{*}})}~.\nonumber \end{eqnarray}
This kernel is the inverse of the Feynman propagator (\ref{eq:nfprop})\[
\int d^{4}z\sqrt{{-g(z)}}K(x,z)\tilde{G}_{F}^{\alpha}(z,y)=-\frac{\delta^{4}(x-y)}{\sqrt{{-g(x)}}}~.\]
It is possible to make a field redefinition to write the kinetic term
in local form, but then the potential term becomes non-local. It is
also worth noting that the amplitudes (\ref{eq:tord}) cannot be interpreted
simply as amplitudes in a squeezed state background, contrary to \cite{Einhorn:2003xb}.
This would require the $U$ operators to be commuted past the time-ordering
symbol, so that one could define an asymptotic state $U|E\rangle$
in the infinite past. However this step cannot be made due to the
non-locality of the theory, as one can easily check using the explicit
mode expansions of the amplitudes. Summing up, this approach differs
from the squeezed state approach described in section \ref{sub:squeezed},
due to the different time-ordering prescription, which in turn leads
to a different non-local kinetic term.

\subsection{Algebra of observables}

Local sources couple directly to $\tilde{\phi}$ and likewise interactions
are local (\ref{eq:locac}) \cite{Goldstein:2003ut}. If we are interested
in the scalar field theory with possible local scalar couplings of
other fields to $\tilde{\phi}$, then the set of observables will
be built out of local products of $\tilde{\phi}$ and derivatives.
As shown in \cite{Allen:1985ux}, the commutator algebra of the $\tilde{\phi}$
is actually independent of $\alpha$, and so vanishes at spacelike
separations. The same will therefore be true of local products of
the $\tilde{\phi}$. Apparently then the pure scalar field theory
can give rise to a self-consistent set of probability amplitudes in
this approach, which does not allow faster than light signaling. This
provides us with a posteriori justification for taking the single-component
real-time contour leading to (\ref{eq:tord}).

However gravity couples locally to the stress-energy tensor\begin{eqnarray*}
T_{\mu\nu}(x) & = & \frac{2}{\sqrt{-g(x)}}\frac{\delta S[\tilde{\phi}]}{\delta g^{\mu\nu}(x)}\\
 & = & a\left(\tilde{\phi}_{;\mu}(x)\tilde{\phi}_{;\nu}(x)-\frac{1}{2}g_{\mu\nu}(x)g^{\rho\sigma}(x)\tilde{\phi}_{;\rho}(x)\tilde{\phi}_{;\sigma}(x)+\right.\\
 &  & \left.\frac{1}{2}m^{2}\left(\tilde{\phi}(x)\right)^{2}g_{\mu\nu}(x)\right)+\left(V(\tilde{\phi}(x))+j(x)\tilde{\phi}(x)\right)g_{\mu\nu}(x)+\\
 &  & b\left(\frac{1}{2}\tilde{\phi}_{;\mu}(x)\tilde{\phi}_{;\nu}(\bar{x})+\frac{1}{2}\tilde{\phi}_{;\mu}(\bar{x})\tilde{\phi}_{;\nu}(x)-\frac{1}{2}g_{\mu\nu}(x)g^{\rho\sigma}(x)\tilde{\phi}_{;\rho}(x)\tilde{\phi}_{;\sigma}(\bar{x})+\frac{1}{2}g_{\mu\nu}(x)m^{2}\tilde{\phi}(x)\tilde{\phi}(\bar{x})\right)\end{eqnarray*}
which is non-local in the $\tilde{\phi}$'s due to the non-local kinetic
term (\ref{eq:kinetic}). The commutator of $T^{\mu\nu}$ with a local
product of $\tilde{\phi}$ can therefore be non-vanishing at spacelike
separations. Therefore once the scalar field is coupled to gravity,
the locality of the observables is spoiled and faster than light signaling
becomes possible. This should be taken as a sign that the theory is
non-perturbatively ill-defined once coupled to gravity. Once faster
than light signaling is possible, it should be possible to consider
processes analogous to closed timelike curves, which typically lead
to uncontrollable quantum backreaction \cite{Hawking:1992nk}. We
emphasize this is not acausality at Planck scale separations, but
macroscopic acausality induced by propagation of the massless graviton.
Even if these terms appear with tiny coefficients (as they would based
on the arguments of \cite{Goldstein:2002fc,Alberghi:2003br}), there
is no known theoretical framework for handling such processes.

One could try to fix this problem by placing the graviton itself in
an $\alpha$-state, by instead demanding a linear combination of $g_{\mu\nu}(x)$
and $g_{\mu\nu}(\bar{x})$ couple locally to $T^{\mu\nu}$ . As with
the $\tilde{\phi}$ field, the new graviton will then have a non-local
kinetic term. This may well work at the linearized level around a
fixed de Sitter background, but once we include gravitational interactions
and proceed to write down a diffeomorphism invariant action, one will
run into problems. For the pure scalar field theory to work it was
important that interactions were local, despite the non-local kinetic
term. However, if we start with the non-local gravitational kinetic
term and add interactions order by order in Newton's constant to achieve
diffeomorphism invariance, we will induce non-local gravitational
interaction terms. Again it seems impossible to avoid problems with
faster than light signaling. 

Furthermore the anti-podal symmetry is a special feature of de Sitter
space that will not generalize in a background independent way. One
could try to define the theory on the Lorentzian continuation of $\mathbb{RP}^{4}$,
making the identification $x\sim\bar{x}$. Here the gravitational
action takes the conventional Einstein-Hilbert form, but it is not
clear how to make sense of physics on such a spacetime. One could
take as a fundamental region the inflationary patch\[
ds^{2}=1/\eta^{2}\left(-d\eta^{2}+d\vec{x}^{2}\right)~,\]
and treat $\eta$ as the global time coordinate. However as you extend
a time-like geodesic to the past, you eventually hit the line $\eta=-\infty$
in finite proper time, and then begin moving forward in time at a
different spatial position. Therefore the spacetime is not time orientable.
As described in \cite{Parikh:2002py}, this implies global quantization
of a free scalar field on this spacetime is not possible. In \cite{Parikh:2002py}
it is argued scalar field quantization within a single static patch
can be done self-consistently. However without a global quantization
method, one must go well beyond the conventional framework of semi-classical
quantum gravity to make sense of the coupling of such a system to
gravity.

For the Bunch-Davies vacuum all these problems are avoided, because
the kinetic term is local for $e^{\alpha}=0$. We conclude that the
Bunch-Davies vacuum is the only de Sitter invariant vacuum state that
yields a consistent conventional perturbative quantum field theory
when coupled to gravity.

\section{Conclusions}

We have seen that from the current theoretical standpoint, the $\alpha$-vacua
are in general inconsistent. The most straightforward treatment as
a squeezed state leads to pinch singularities which render the perturbative
expansion ill-defined. Another approach derived from imaginary time
methods leads to a well-defined perturbative expansion, however the
theory becomes non-local once coupled to gravity. We stress this non-locality
is over macroscopic scales due to the fact it is induced by the massless
graviton, so does not have a local effective description even at arbitrarily
low energies. Conventional wisdom then suggestions the vacua cannot
be consistently coupled to gravity at the non-perturbative level.
Hopefully these results serve to pin-point the problems with the so-called
$\alpha$-vacua, and establish the Bunch-Davies vacuum as the unique
de Sitter invariant vacuum state that survives coupling to gravity.

These results have a number of important implications for trans-Planckian
effects during inflation. Certain classes of trans-Plankian effects
can be modelled as an $\alpha$-vacua with an explicit ultra-violet
cutoff as advocated in \cite{Goldstein:2002fc,Danielsson:2002kx}.
In these models it is presumed unknown ultra-violet physics place
modes in an $\alpha$-vacuum below some proper cutoff wavenumber.
The present results indicate it is unlikely this unknown ultraviolet
physics can be described by a local perturbative effective field theory.
Within the context of local effective field theory, one can still
fine-tune the initial state so that it gives rise to unusual effects
at the end of inflation. However, we now can convincingly argue that
generic perturbations will inflate away and the unique de Sitter invariant
Bunch-Davies state will be left behind %
\footnote{One might hope that inflation was sufficiently short for certain pertubations
to survive - see for example \cite{Kaloper:2003nv}.%
}. Up to fine tuning issues, the influence of high energy physics on
inflation can then be captured by a local low-energy effective action
analysis around the Bunch-Davies state, which leads to the conclusion
that high energy physics corrections to the cosmic microwave background
will typically be beyond the cosmic variance limits \cite{Kaloper:2002cs,Kaloper:2002uj}
(notwithstanding some loop-holes \cite{Burgess:2003zw,Burgess:2002ub}).
One can also view these results as highlighting the type of modification
of conventional gravity needed to make sense of a variety of proposed
trans-Planckian effects. Finally, it should also be noted our results
do not apply to trans-Planckian corrections to inflation that do not
asymptote to a de Sitter $\alpha$-vacuum (see for example \cite{Tsujikawa:2003gh}).

\begin{acknowledgments}
We thank R. Brandenberger for helpful comments. D.L. thanks the Departamento
de F\'{i}sica, CINVESTAV for hospitality. This research is supported
in part by DOE grant DE-FE0291ER40688-Task A and US--Israel Bi-national
Science Foundation grant \#2000359.

\bibliographystyle{apsrev}
\clearpage\addcontentsline{toc}{chapter}{\bibname}\bibliography{desitterqft}
\end{acknowledgments}

\end{document}